\documentclass[fleqn,twoside]{article}
\usepackage{espcrc2}

\usepackage{graphicx}
\usepackage[figuresright]{rotating}

\newcommand{\AmS}{{\protect\the\textfont2
  A\kern-.1667em\lower.5ex\hbox{M}\kern-.125emS}}

\title{Experimental test of the probability density function of true value 
of Poisson distribution parameter by single observation of number of events}

\author{S.I. Bityukov\address[IHEP]{Institute for high energy physics, 
142281 Protvino, Russia}%
\thanks{The participation in the Workshop is supported by the 
Organizing Committee of ACAT'03.}, 
V.A. Medvedev\addressmark[IHEP], 
V.V. Smirnova\addressmark[IHEP], 
Yu.V. Zernii\address[MGAPI]{Moscow State Academy of Instrument Engineering
and Computer Science, Moscow, Russia}}

\begin{document}

\begin{abstract}
The empirical probability density function for the conditional distribution
of the true value of Poisson distribution parameter on one measurement
is constructed by computer experiment. The analysis of the obtained
distributions  confirms that these distributions are gamma-distributions.
\vspace{1pc}
\end{abstract}

\maketitle

\section{INTRODUCTION}

Let us consider the Gamma-distribution with probability density

\begin{equation}
g_x(\beta,\alpha) = \displaystyle 
\frac{x^{\alpha-1}e^{-\frac{x}{\beta}}}{\beta^{\alpha}\Gamma(\alpha)}.   
\end{equation}

At change of standard designations of Gamma-distribution 
$\displaystyle \frac{1}{\beta}$, $\alpha$ and $x$ for  
$a$, $n+1$ and $\lambda$ we get the following formula for probability density 
of Gamma-distribution 

\begin{equation}
g_n(a,\lambda) = \displaystyle 
\frac{a^{n+1}}{\Gamma(n+1)} e^{-a\lambda} \lambda^{n},   
\end{equation}

\noindent
where $a$ is a scale parameter and $n + 1 > 0$ is a shape parameter. 
Suppose $a = 1$, then the probability density of Gamma-distribution 
$\Gamma_{1,n+1}$ looks like Poisson distribution of probabilities:

\begin{equation}
g_n(\lambda) = \displaystyle \frac{\lambda^n}{n!} e^{-\lambda},~ 
\lambda > 0,~n > -1.  
\end{equation}

Let the probability of observing $n$ events in the experiment 
be described by a Poisson distribution with parameter $\lambda$, i.e.

\begin{equation}
f(n; \lambda)  = \frac{{\lambda}^n}{n!} e^{-\lambda}.
\end{equation}

As it follows from the article~\cite{1} (see also~\cite{2}) 
and is clearly seen from the analysis of identity~\cite{3}

\begin{equation}
\displaystyle
\sum_{k = n + 1}^{\infty}{f(k; \lambda_1)} +
\int_{\lambda_1}^{\lambda_2}{g_{n}(\lambda) d\lambda} + 
\sum_{k = 0}^{n}{f(k;\lambda_2)} = 1,~
\end{equation}

\noindent
i.e.

\begin{center}
$\displaystyle
\sum_{k = n + 1}^{\infty}{\frac{\lambda_1^ke^{-\lambda_1}}{k!}} +
\int_{\lambda_1}^{\lambda_2}
{\frac{\lambda^{n}e^{-\lambda}}{n!}d\lambda}
+ \sum_{k = 0}^{n}{\frac{\lambda_2^ke^{-\lambda_2}}{k!}} = 1~$
\end{center} 

\noindent
for any $\lambda_1 \ge 0$ and $\lambda_2 \ge 0$, that at one measurement
of the number of events $n$ (in our case it is the number of casual events
appearing in some system for certain period of time) which appear according
to Poisson distribution, the parameter value of this distribution is
described by Gamma-distribution $\Gamma_{1,1+n}$ with mean, mode, and variance 
$n+1,~n$, and $n+1$, respectively. In other words conditional distribution
of the probability of true value of parameter of Poisson distribution
is a Gamma-distribution $\Gamma_{1,1+n}$ on condition that the measured
value of the number of events is equal to $n$.

It means that the value $n$ corresponds to the most probable   
value of parameter of Poisson distribution, and the mean value of the  
number of events, appearing in Poisson flow in the fixed time interval,
must correspond to the magnitude $n+1$,   
i.e. the estimation of parameter of Poisson distribution at one observation 
is displaced for 1 from the measured value of the number of events. 
The equation~(5) in the considered case allows to mix Bayesian and 
frequentist probabilities.
 
As a result, we can easily construct the confidence 
intervals, take into account systematics and statistical uncertainties of
measurements at statistical conclusions about the quality of planned 
experiments, estimate the value of the parameter of Poisson distribution 
by several observations~\cite{3,4}.

Nevertheless there are works in which the approaches based on other 
assumptions of distribution of true value of parameter of Poisson distribution 
in presence of its estimation on a single measurement, for example~\cite{5}.
Also the works using Monte Carlo methods for construction of 
confidence intervals and for estimations of Type I and Type II errors 
in the hypotheses testing have recently appeared~\cite{6,7} (see, also,
\cite{9}).
Therefore the experimental test with the purpose to confirm,  
that the true value of parameter of Poisson distribution at
single observation has density of probability of Gamma-distribution,
and with the purpose to check up applicability of Monte Carlo methods 
to such tasks was carried out.

The structure of the paper is the following. In the next section the 
arrangement of computer experiment is considered,
in the third section - the statistical analysis of the results is given,
and the last section contains concluding remarks. 

\section{The arrangement of measurements}

From the identity (5) follows, that any prior, except the uniform, on value of 
parameter of Poisson distribution in distribution of true
value of this parameter at presence of the measured estimation $n$ 
is excluded by existence of the boundary conditions determined by the 
appropriate sums of Poisson distributions (see Eq.(5)).
Therefore we carried out the uniform scanning in parameter of Poisson
distribution with step 0.1 from value $\lambda=0.1$ up to value
$\lambda=20$, playing the Poisson distribution 30000 trials for each value 
$\lambda$ (Fig.1) with the using of function RNPSSN~\cite{8}.

\begin{figure}[htb]
\centering \includegraphics[height=0.75\hsize]{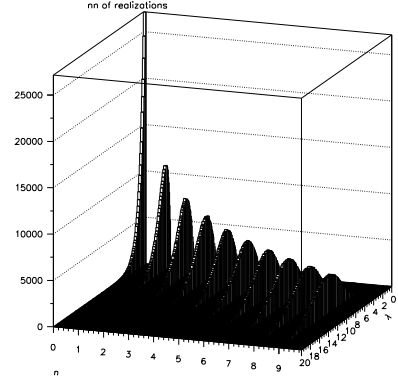}
\caption{\small Amount of occurrences of $n$ in the interval from 0 
up to 9. Scanning in parameter of the Poisson distribution
(30000 trials at each value of parameter $\lambda$) was carried out with 
step 0.1 in the interval of $\lambda$ from 0.1 up to 20 }
\label{fig:1}
\end{figure}

After scanning for each value of number of the dropped out events $n$ 
the empirical density of probability of true value of parameter of Poisson
distribution to be $\lambda$ if the observation is equal $n$ was obtained.

\section{The analysis of results }

In Fig.2 the distribution (a), obtained at scanning (the playing of Poisson 
distribution with consequentive increase of the parameter $\lambda$ after 
each series of trials with the fixed value of the parameter) in parameter  
$\lambda$ with the selection of number of the dropped out events $n=6$,
and distribution (b) of the casual value, having $\Gamma_{1,7}$ distribution
of appropriate area, calculated by the formula, are shown.
One can see that the average value of parameter $\lambda \approx 7$.
It means, the number of observed events is displaced for one unity by the
estimation of the mean value of Poisson distribution parameter and correspond
to the most probable value (the mean value has bias).

\begin{figure}[htpb]
\centering \includegraphics[height=0.75\hsize]{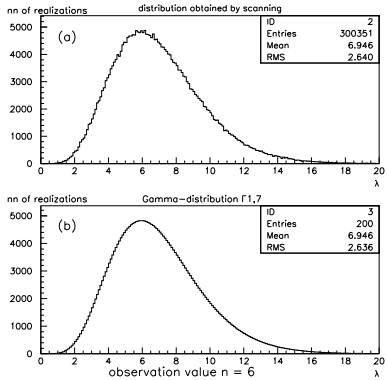}
\caption{\small Distributions of occurrences of value $n=6$ depending on 
value of parameter $\lambda$. The distribution (a) is obtained at Monte Carlo 
scanning in parameter $\lambda$. The distribution (b) is obtained by direct 
construction of Gamma-distribution $\Gamma_{1,7}$} 
\label{fig:2}
\end{figure}

The same distributions obtained by the selection of number of dropped out 
events $n=0$ (Fig.3) and $n=8$ (Fig.4) superimposed on each other
in logarithmic scale also are shown.

\begin{figure}[htpb]
\centering \includegraphics[height=0.75\hsize]{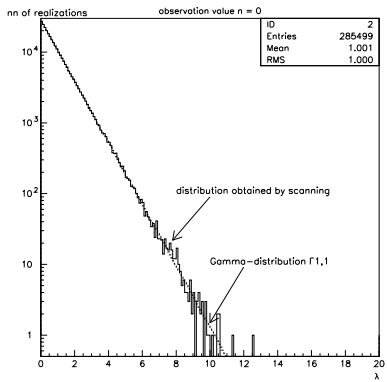}
\caption{\small Distributions of occurrences of value $n=0$ depending on 
value of parameter $\lambda$. The distribution obtained at Monte Carlo 
scanning in parameter $\lambda$ is superimposed on the distribution 
obtained by direct construction of Gamma-distribution $\Gamma_{1,1}$} 
\label{fig:3}
\end{figure}

\begin{figure}[htpb]
\centering \includegraphics[height=0.75\hsize]{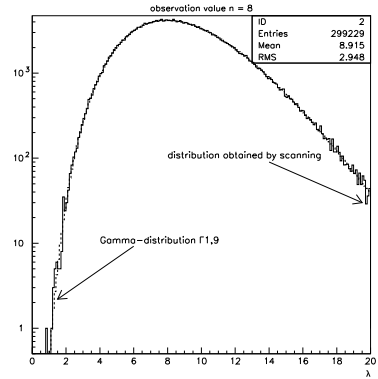}
\caption{\small Distributions of occurrences of value $n=8$ depending on 
value of parameter $\lambda$. The distribution obtained at Monte Carlo 
scanning in parameter $\lambda$ is superimposed on the distribution 
obtained by direct construction of Gamma-distribution $\Gamma_{1,9}$} 
\label{fig:4}
\end{figure}

In Tab.1 the values of probabilities of compatibility of the 
empirical distribution, obtained by Monte Carlo by the scanning in parameter 
$\lambda$, and the appropriate Gamma-distribution for values $n$ 
from 0 up to 9 are presented.
The calculations are based on the Kolmogorov Test (the function HDIFF of 
the package HBOOK~\cite{8}).
The authors of a package as criterion of coincidence of two distributions 
recommend to use the requirement of probability value of compatibility
more than 0.05. In Fig.4 the least conterminous distributions are given.

\begin{table}[htb]
\caption{The probability of compatibility}
\label{table:1}
\begin{tabular}{|r|r||r|r|}
\hline
$n$ &  probability &  $n$ &  probability \\ 
\hline
 0  &  1.000000  & 5  &  0.999084  \\ 
 1  &  0.999646  & 6  &  0.999986  \\
 2  &  0.992521  & 7  &  0.999892  \\
 3  &  0.999986  & 8  &  0.752075  \\
 4  &  0.999969  & 9  &  0.974236  \\
\hline
\end{tabular}\\[2pt]
\end{table}

Thus, the obtained results do not contradict the statement that conditional 
distribution of true value of parameter of Poisson distribution at
single measurement is obeyed to a Gamma-distribution.

\section{Conclusion}

In the report Monte Carlo experiment on the check of the   
statement, that true value of parameter of Poisson distribution  at an 
estimation of this parameter on one observation $n$ has probability
density of Gamma-distribution $\Gamma_{1.n+1}$, is carried out.
The obtained results confirm the conclusions of the papers~\cite{3,4} about 
a kind of conditional distribution of true value of parameter of 
Poisson distribution at single observation.

Note, that the given results also specify the applicability of 
Monte Carlo method for construction of conditional distribution of
the true value of parameters of various distributions.

\begin{center}
{\bf Acknowledgment}
\end{center}

The authors thank N.V.~Krasnikov, V.F.~Obraztsov, V.A.~Petukhov and 
M.N.~Ukhanov for support of the given work. The authors also are grateful to 
S.S.~Bityukov for fruitful discussions and constructive criticism. 
S.B. thank Toshiaki Kaneko and Fukuko Yuasa.
The authors wish to thank E.A.~Medvedeva for help in preparing the paper.
This work has been supported by grant RFBR 03-02-16933.

\newpage

\begin{center}
                {\Large  Appendix}
\end{center}

Let us consider the famous equality~\cite{1,2} in form as written 
in~\cite{3}

\begin{equation}
\displaystyle \sum_{i=n+1}^{\infty} f(i;\mu_1) +
    \int_{\mu_1}^{\mu_2}{g(\mu;n)d\mu} + \sum_{i=0}^{n} f(i;\mu_2) = 1, 
\end{equation}

where $\mu_1 \ge 0$, $\mu_2 \ge 0$, \\
$\displaystyle f(k;\mu) = g(\mu;k)= \frac{\mu^k e^{-\mu}}{k!}$, 
$k=1,2, \dots .$ 

\bigskip

Let us suppose that $g(\mu;k)$ is the probability density of parameter of the 
Poisson distribution to be $\mu$ if number of observed events is equal to k.
It is a conditional probability density. As shown above (Eq.3) 
the $g(\mu;k)$ is the density of Gamma-distribution by definition. 

\bigskip

On other side: if $g(\mu;k)$ is not equal to this probability density and
the true probability density of the Poisson parameter is the other 
function $h(\mu;k)$ (continuous or with set of points of discontinuity 
by measure 0) then there takes place another identity

\newpage

\begin{equation} 
\displaystyle \sum_{i=n+1}^{\infty} f(i;\mu_1) +
    \int_{\mu_1}^{\mu_2}{h(\mu;n)d\mu} + \sum_{i=0}^{n} f(i;\mu_2) = 1, 
\end{equation}

This identity is correct for any $\mu_1 \ge 0$ and $\mu_2~\ge~0$.

\bigskip

If we subtract Eq.7 from Eq.6 then we have

\begin{equation}
 \int_{\mu_1}^{\mu_2}{(g(\mu;n) - h(\mu;n))d\mu}  = 0.
\end{equation}

We can choose the $\mu_1$ and $\mu_2$ by the arbitrary way. Let us make
this choice so that $g(\mu;n)$ is not equal $h(\mu;n)$ in the interval
$(\mu_1,\mu_2)$ and, for example, $g(\mu;n) > h(\mu;n)$ and
$\mu_2 > \mu_1$. In this case we have 

\begin{equation}
\int_{\mu_1}^{\mu_2}{(g(\mu;n) - h(\mu;n))d\mu} > 0
\end{equation}

\noindent
and as a result we have contradiction.  The identity (6) does not leave 
a place for any prior except uniform.

\end{document}